\documentclass[conference,10pt, letterpaper]{IEEEtran}
\usepackage{graphicx}
\usepackage{amssymb}
\usepackage{amsmath}
\usepackage{amsfonts}
\usepackage{cite}
\usepackage{url}
\usepackage{enumerate}
\usepackage{epstopdf}
\usepackage{dsfont}
\usepackage{suffix}
\usepackage{steinmetz}
\usepackage{amsthm}
\usepackage{color}
\usepackage{algpseudocode}
\usepackage{algorithm}
\usepackage{balance}
\usepackage[mathscr]{euscript}

\DeclareSymbolFont{sfletters}{OML}{cmbrm}{m}{s1}
\DeclareMathSymbol{\sdelta}{\mathord}{sfletters}{"0E}

\DeclareMathSymbol{\seta}{\mathord}{sfletters}{"11}

\allowdisplaybreaks[1]

\makeatletter
\def\footnoterule{\relax%
  \kern-5pt
  \hbox to \columnwidth{\hfill\vrule width \columnwidth height 0.4pt\hfill}
  \kern4.6pt}
\makeatother

\begin{document}
%\title{Noncoherent Detection and Channel Coding for Scatter Radio Sensor Networking} 
%\title{Noncoherent Short Packet Receivers for Scatter Radio Sensor Networks}
\title{Nonlinear Energy Harvesting Models \\in Wireless Information and Power Transfer}
%  Nonlinear Energy Harvesting Models in Wireless Information and Power Transfer

\author{
        \hspace{4cm}\\[-0.6cm]
        Panos N. Alevizos,
        Georgios Vougioukas, and 
        Aggelos Bletsas \\
        School of ECE, Technical Univ. of Crete, Kounoupidiana Campus, Greece 73100\\
         {\small e-mail: {\{palevizos, gevougioukas\}@isc.tuc.gr}, {aggelos@telecom.tuc.gr}}
\thanks{This work was implemented in the context of Elidek...
}
\thanks{Authors are with School of Electrical and Computer Engineering (ECE), Technical University of Crete, Chania 73100, Greece (e-mail: 
{\tt \{gevougioukas, palevizos\}@isc.tuc.gr}, {\tt aggelos@telecom.tuc.gr})
}
}

%Nonlinear Energy Harvesting Models in Wireless Information and Power Transfer
\maketitle

\begin{abstract}
This work  compares different linear and nonlinear RF energy harvesting models, including limited or unlimited sensitivity, for  simultaneous wireless information 
and power transfer (SWIPT). The  probability of successful SWIPT reception under a family of RF harvesting models is rigorously quantified, 
using state-of-the-art rectifiers in the context of commercial RFIDs. A significant portion of SWIPT literature uses oversimplified models 
that do not account for limited sensitivity or nonlinearity of the underlying harvesting circuitry. This work demonstrates that communications signals are not always appropriate for simultaneous energy transfer and concludes that for
practical SWIPT studies, the inherent non-ideal characteristics of the harvester should be carefully  taken into account; specific harvester's modeling methodology is also offered.
\end{abstract}

% \begin{IEEEkeywords}
% RF Energy Harvesting, SWIPT.
% \end{IEEEkeywords}

%\begin{IEEEkeywords}
%Rician channels
%\end{IEEEkeywords}
\IEEEpeerreviewmaketitle

%\vspace{-0.06 in}
\section{Introduction}

Intense research has been devoted the last years on simultaneous wireless information and power transfer (SWIPT). 
The main concept in far field SWIPT systems is the exploitation of the communication signals for radio frequency (RF)
energy harvesting, typically with rectennas, i.e., antenna and rectifier(s). The latter perform the required RF-to-DC conversion,
including one (or more) diode(s). The main problem in far field RF energy harvesting is the limited sensitivity of the circuit,
currently in the order of $-35$ dBm to $-25$ dBm, with slow improvement by a factor of $2$ every approximately $5$ years \cite{Du:16}.
Such power levels below which energy transfer cannot be performed, are orders of magnitude higher than current communications circuits
sensitivity, which may reach values as low as $-130$ dBm to $-80$ dBm, depending on bandwidth. Thus, signals appropriate for communications 
may not be \emph{simultaneously} suitable for energy transfer \cite{Ale_PhD:17}, \cite{AlBl:18}. 

Another major issue in the SWIPT literature is the adoption of oversimplified RF harvesting models, which either exhibit a linear relationship between input RF and output harvested power or assume unlimited sensitivity. Rectennas, due to the presence of diode(s), exhibit a highly nonlinear behavior, with limited sensitivity, due to the need for bias.  Despite the vast amount of literature in the wireless communications theory community that adheres to the above assumptions, exceptions have only recently started to emerge; for example, work in \cite{CleBaz:16,Cler:18} utilized convex optimization techniques to optimize the  parameters
of multi-tone waveforms, which improve RF harvesting efficiency compared to single-tone, while taking into account the nonlinearity of the rectifier. Other nonlinear RF harvesting models have  been recently proposed, which however miss the limited sensitivity issue and will be discussed subsequently.

Therefore, there is a strong need to evaluate different RF harvesting models, taking into account \emph{both} harvesting sensitivity \emph{and} nonlinearity, as well as facts from the relevant microwave literature.   Radio frequency identification (RFID) technology is the most prominent example of SWIPT, with significant prior art, as well as commercial interest. This work compares different linear and nonlinear energy harvesting models for SWIPT, taking also into account limited or unlimited sensitivity; comparisons are performed based on real, state-of-the-art rectifiers \cite{AsDaBle:16} in RFID, using backscatter communications. It is found that neglecting harvester's nonlinearity and limited sensitivity may offer misleading results. 

%the way incident RF power is exploited for powering the tag

%Radio frequency identification (RFID) is a concrete, industrial simultaneous wireless information and power transfer (SWIPT) example and dates back to the early 1940's \cite{St:48}. It is based on backscatter radio, i.e., communications by means of reflection, due to its extraordinary ultra-low power nature, recently exploited in digital sensor networking \cite{VaBlLe:08}, analog sensor networking \cite{KaKiTouKoKoBl:14, DaAsKaBl:16}, and IoT \cite{AlBlKar:17, Ale_PhD:17}. Thus, it is interesting to examine such paradigm, under the prism of recent findings in the SWIPT community.

\section{Signal Model}
\label{sec:signal_model}

\begin{figure}[!t]
 \centering
  \includegraphics[width=0.95\columnwidth]{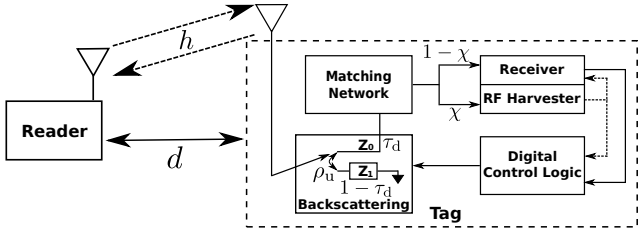}
  \caption{Monostatic backscatter architecture consisting of a reader
  and a passive (i.e., batteryless) RFID tag. Reader acts as carrier emitter,
  as well as receiver of tag reflected/backscattered information.}
 \label{fig:reader_tag_RFID}
 \end{figure}

Backscatter  radio/RFID technology is the most prominent example  of SWIPT. A monostatic, single-antenna reader topology is  examined with reader and tag,  depicted in Fig.~\ref{fig:reader_tag_RFID}.
In that case, the illuminating carrier emitter and the receiver of the tag-backscattered signal is the same, full-duplex unit, a.k.a.  the reader;
the latter is equipped with a single antenna serving both reception and transmission, using an appropriate duplexer, the circulator. 
Thus, path-loss and small-scale fading are the same for both reader-to-tag (downlink) and tag-to-reader (uplink) links. Both links are 
subject to large-scale fading, where the path-gain at tag-to-reader distance $d$ is given by:
\begin{equation}
\mathsf{L} \equiv \mathsf{L}( {d}) =  
\left(\frac{\lambda}{4 \pi d_0}\right)^2 \left(\frac{d_0}{   {d}}\right)^{\nu},
\label{eq:path_loss}
\end{equation}
where   $d_0$ is a reference distance (assumed unit thereinafter), 
$\lambda$ is the wavelength and $\nu $ is the path loss exponent.

Flat fading is assumed due to relatively small communication bandwidth. Thus, small-scale fading coefficient, for both downlink and uplink is given by $h = a \mathsf{e}^{-\mathsf{j} \phi}$. Due to potential strong line-of-sight (LoS), Nakagami small-scale fading is 
assumed with ${\mathbb{E}} \!\left[ a^2 \right]   =1$ and Nakagami parameter $\mathtt{M} \geq \frac{1}{2}$ \cite[p.~79]{Goldsmith:05}. The special cases of Rayleigh fading  and no fading ($a = 1$) are obtained for $\mathtt{M} = 1$ and $\mathtt{M} = \infty$, respectively.

%\color{blue}Reader emits a carrier with, passband representation:
%\begin{equation}
%\mathsf{c}_{\rm R}(t) = \sqrt{2 \, \mathtt{P}_{\rm R} }\,\Re \{ \mathsf{e}^{ \, \mathsf{j} 2 \pi F_{\rm c}  t} \},
%\end{equation}   
%where $ \mathtt{P}_{\rm R}$ is the reader's transmit power and $F_{\rm c}$ carriers' center frequency. 

Assuming the reader emits an unmodulated carrier with transmit power $\mathtt{P}_{\rm R}$ and frequency $F_{\rm c}$, the impinged signal at the tag signal can be expressed as follows:
\begin{equation}
 \mathsf{c}_{\rm T}(t)  = \sqrt{2 \, \mathsf{L} \,\mathtt{P}_{\rm R} }\,\Re \{ h \, \mathsf{e}^{ \, \mathsf{j} 2 \pi F_{\rm c}  t} \}.
\label{eq:tag_received_signal}
\end{equation}
The received power at the tag is then given by:
\begin{equation}
 P_{\rm in} =  \mathsf{L} \,\mathtt{P}_{\rm R} \,|h|^2 = \mathsf{L} \,\mathtt{P}_{\rm R} \, a^2.
 \label{eq:input_power}
\end{equation}
According to the above, $P_{\rm in}$ follows Gamma distribution (${\mathbb{E}} \!\left[ a^2 \right]   =1$):
$
 \mathsf{f}_{P_{\rm in} }(x)   =  \left( \frac{\mathtt{M}}{\mathsf{L}\, \mathtt{P}_{\rm R}}\right)^{\mathtt{M}} \,
 \frac{x^{ \mathtt{M}- 1}}{\mathsf{\Gamma}(\mathtt{M})} \, \mathsf{e}^{-\frac{\mathtt{M}}{\mathsf{L} \, \mathtt{P}_{\rm R}} x}, ~x \geq 0, 
\label{eq:gamma_distr_Pin}
$
where $\left( \mathtt{M} , \frac{\mathsf{L}  \,  \mathtt{P}_{\rm R}}{\mathtt{M}}\right)$ the shape and scale parameter, respectively, and $\mathsf{\Gamma}(x)= \int_{0}^{\infty}t^{x-1} \mathsf{e}^{-t} \mathsf{d}t$ is the Gamma function.

\section{RFID Tag Operation}
\label{sec:tag_operation}
The RFID tag does not include any power-demanding signal
conditioning units, e.g., amplifiers, mixers or oscillators (Fig.~\ref{fig:reader_tag_RFID}). Instead, communication is achieved by varying the reflection coefficient between tag antenna and its termination loads, using a RF switch. 
Binary modulation is achieved with two different reflection coefficients (i.e., two different termination
loads $Z_0$, $Z_1$). This operation results to modulation of tag information on top of the reader
illuminating signal, reflected (from the tag) back to the reader, in an ultra low-power fashion.

\subsection{RF Harvesting \& Tag Powering}
\label{subsec:tag_harvesting}
In order for the RFID tag to operate, power must be harvested from the impinged, reader-generated signal.
Input power must be above the tag harvester sensitivity $\mathtt{P}_{\rm sen}$, i.e.,
$ P_{\rm in} > \mathtt{P}_{\rm sen}$. $\mathtt{P}_{\rm sen}$ is a crucial parameter in backscatter 
communication with passive tags, due to the fact that state-of-the-art, far field RF harvesters offer limited sensitivity.

Work in \cite{AlBl:18} established that  a high-order polynomial in the dBm scale can be
safely considered as ground truth model for harvesting efficiency function; thus, harvested   power can be modeled as a function of input power $x$  as follows:
\begin{align}
\mathsf{p}(x) = \begin{cases}
                0,   & x \in [0,\mathtt{P}_{\rm sen}) \\
 \big ( w_0 + \sum_{i=1}^{W}  w_i  (10 \, \mathsf{log}_{10}(x) )^i \big) \cdot x , & x\in [\mathtt{P}_{\rm sen},\mathtt{P}_{\rm sat} ], \\
 \mathsf{p}(\mathtt{P}_{\rm sat}),&x \geq \mathtt{P}_{\rm sat},
 \end{cases}
\label{eq:harvested_power_realistic_function}
\end{align} 
where $x$ and $\mathsf{p}(x)$ take values in mWatt,  while  the quantity $\big ( w_0 + \sum_{i=1}^{W}  w_i  (10 \, \mathsf{log}_{10}(x) )^i \big)$  
is the harvesting efficiency function,
with  $W$ being  the degree of the polynomial  and   $\{w_i\}_{i=0}^W$   the corresponding coefficients. 
For the analysis below we assume that function $\mathsf{p}(x)$ is continuous and increasing in $[\mathtt{P}_{\rm sen},\mathtt{P}_{\rm sat} ]$. 
As shown in \cite{AlBl:18},  the parameters $\{w_i\}_{i=0}^W$ in Eq.~\eqref{eq:harvested_power_realistic_function} can be obtained directly from harvesters' data using standard  convex optimization fitting methods.

Several models have been proposed in order for the harvested power to be mathematically described. These models are summarized below:
\subsubsection{Linear Model (L)}
\label{subsubsec:linear_model}
Single parameter model, where the harvested power can be expressed as $\mathsf{p}_{1}(x) \triangleq \eta_{\rm L} x,~x\geq 0$. This is the most utilized model in SWIPT literature, it's linear and does not account for harvesters' sensitivity.
\subsubsection{Constant Linear (CL)}
\label{subsubsec:constantLinear_model}
Linear model with the addition of taking into account the sensitivity of the harvester. According to that model, harvested power is expressed 
as $\mathsf{p}_{2}\! \left(  x  \right) = \eta_{\rm CL} \cdot (x- \mathtt{P}_{\rm sen})$ for $x \in [\mathtt{P}_{\rm sen}, \infty)$ and 
zero in the rest of its domain; $\eta_{\rm CL}$ is the constant harvesting efficiency.
\subsubsection{Nonlinear Normalized Sigmoid}
\label{subsubsec:nonl_sigmoid}
The model was proposed in \cite{BoNgZlSc:15} and assumes $\mathtt{P}_{\rm sen} = 0$, i.e., it does not account for harvesters' sensitivity. The harvested power is expressed as:
\begin{equation}
    {\mathsf{p}}_{3}\! \left(  x  \right) \triangleq 
\frac{\frac{\mathtt{c}_0}{1 + \mathsf{exp} \left(-\mathtt{a}_0(x - \mathtt{b}_0) \right)} - 
\frac{\mathtt{c}_0}{1 + \mathsf{exp}  \left( \mathtt{a}_0 \, \mathtt{b}_0 \right) }}{1 - \frac{1}{1 + \mathsf{exp} \left( \mathtt{a}_0 \, \mathtt{b}_0  \right) } }.
\label{eq:harvested_power_normalized_sigmoid}
\end{equation}
The shape of ${\mathsf{p}}_{3}\! \left(  x  \right)$ is determined by three real numbers $\mathtt{a}_0$, $ \mathtt{b}_0$, and $\mathtt{c}_0$. A similar, sigmoid model accounting however for $\mathtt{P}_{\rm sen}$, was proposed in \cite{WaXiHuWu:17}, where the harvested power is modeled as:

$\hspace{-8pt}{\mathsf{p}}_{4}\! \left(  x  \right) \triangleq 
\max \left\{ 
\frac{\mathtt{c}_1}{ \mathsf{exp}  \left( -\mathtt{a}_1 \mathtt{P}_{\rm sen} + \mathtt{b}_1 \right) }
\left(\frac{1 + \mathsf{exp}  \left( -\mathtt{a}_1 \mathtt{P}_{\rm sen} + \mathtt{b}_1 \right)}{1+
\mathsf{exp}  \left( -\mathtt{a}_1 x + \mathtt{b}_1 \right)}  - 1\right), 0  \right\}.
\label{eq:harvested_power_normalized_sigmoid_sens}
$
%which is parameterized by  by three real numbers  $\mathtt{a}_1$, $ \mathtt{b}_1$, and $\mathtt{c}_1$.}
% It is adopted in \cite{BoZlDaNgSch:17}.
\vspace{+8pt}

\subsubsection{Second Order Polynomial}
\label{subsubsec:so_poly}
In \cite{XuOzKeVi:17} a model based on a second degree polynomial in milliWatt domain has been suggested. 
Following that model, harvested power can be expressed as ${\mathsf{p}}_{5}\! \left(  x  \right) \triangleq 
\mathtt{a}_2 \, x^2 + \mathtt{b}_2 \, x + \mathtt{c}_2$. The above model does not account for $\mathtt{P}_{\rm sen}$. In order to encompass the effect of sensitivity, ${\mathsf{p}}_{5}(\cdot)$ can be modified as
\begin{equation}
  {\mathsf{p}}_{6}(x) \triangleq \mathtt{a}_3(x - \mathtt{P}_{\rm sen})^2  + \mathtt{b}_3(x - \mathtt{P}_{\rm sen}).
  \label{eq:harvested_power_second_order_poly_sens}
\end{equation}
The  parameters of the model in Eq.~\eqref{eq:harvested_power_second_order_poly_sens} are $\mathtt{a}_3$, $\mathtt{b}_3$ and $\mathtt{P}_{\rm sen}$.

%(e.g., from an actual RF harvesters' specifications)

\subsubsection{Piecewise Linear Model} 
\label{subsubsec:piecewise_model}
Given a set of $J+1$ data pairs of input power and corresponding harvested power, denoted as $\{q_j\}_{j=0}^{J}$ and
$\{v_j\}_{j=0}^{J}$, respectively, slopes $l_j \triangleq
\frac{v_{j} - v_{j-1}}{q_j - q_{j-1}}$, $j \in [J]$ are defined,
where $[J] \triangleq \{1,2, \ldots, J\}$. Modeling sensitivity and saturation characteristics is done through
points $q_0 = \mathtt{P}_{\rm sen} $ and $q_J = \mathtt{P}_{\rm sat}$. Having those slopes, the harvested power is given by:
\begin{equation}
\mathsf{p}_7(x)
\triangleq  \begin{cases}
0 &  x   \in [0,  q_0], \\
l_j  ( x -  q_{j-1}) + v_{j-1},  &  x \in ( q_{j-1}, q_j]   , ~\forall j\in [J] ,\\
v_J , &  x \in  [ q_J, \infty).
\end{cases}
\label{eq:P_harv_approx_function}
\end{equation}
Function $\mathsf{p}_7(x)$ is defined using $2(J+1)$ real numbers, easily available from harvesters'
specifications; thus, determining $\mathsf{p}_7(x)$ is straightforward, without any tuning.

It should be noted that the last model can potentially model energy harvesting from other sources, other than RF. For instance, if photodiodes are used in order to harvest energy from either ambient or solar light, the proposed model can describe the harvested power, as a function of illuminance (measured in lux). This statement is
based on the nonlinear behavior of the photodiodes (similarly to RF rectification circuits), when used as harvesting elements (for example see work in \cite{GuKleCheYaAMi:09, FoGuKlePhaYaAmi:13}).

%%% HARVESTER FITTING/DIFFERENT MODELS
  \begin{figure}[!t]
 \centering
         \includegraphics[width=0.75\columnwidth]{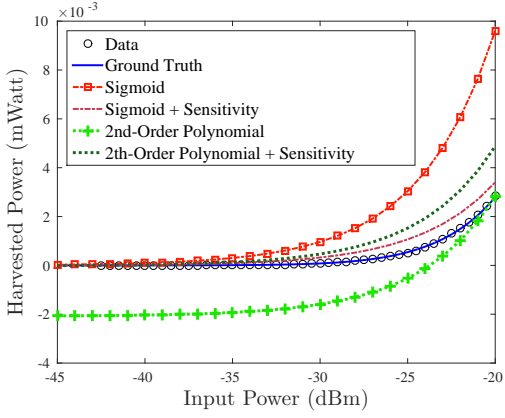}  
  \caption{Harvested power (in milliWatt) versus input power (in dBm) for  the  harvester proposed in \cite{AsDaBle:16} using  nonlinear 
  harvested power function  $\mathsf{p}_n(\cdot)$, $n=3, 4, 5, 6$,
  as  well as for the ground truth model in Eq.~\eqref{eq:harvested_power_realistic_function}.
  Input power range within $[-45,-20]$ dBm.}  
 \label{fig:harv_power_versus_input power}
 \end{figure}

Fig.~\ref{fig:harv_power_versus_input power} illustrates the harvested power (in mWatt) versus input power (in dBm) for the  harvester proposed in \cite{AsDaBle:16} using as
ground truth the specification data; the nonlinear model in Eq.~\eqref{eq:harvested_power_realistic_function} adheres to the data; the rest of the nonlinear harvested power function $\mathsf{p}_n(\cdot)$, $n=3,4,\ldots, 6$, discussed above, are also depicted (using Matlab's fitting toolbox).
Due to strong nonlinearity, the linear models were omitted from the plot.
%  It can be shown that the ground-truth model in~\eqref{eq:harvested_power_realistic_function} is in full accordance with the provided
%  harvested power data of the rectifier.

During normal operation, tags' antenna is terminated at load 
$Z_0$ (absorbing state, see Fig.~\ref{fig:reader_tag_RFID}) for a time fraction of $\tau_{\rm d}$ while for the rest $1 - \tau_{\rm d}$, antenna is connected 
to $Z_1$ (reflection state). Given that the tag is at $Z_0$, a portion $\chi$ of the received power is destined solely for energy harvesting, 
i.e., $\zeta_{\rm har} = \chi \,\tau_{\rm d} \,  \in (0,1)$ percentage of input power is dedicated for RF energy harvesting. The rest 
$(1- \chi )\tau_{\rm d}$, is exploited for downlink communication purposes.

Thus, in order for the tag to operate, the total harvested power $\mathsf{p}( \zeta_{\rm har}  \,  P_{\rm in} )$ must be greater than the tag overall power consumption $\mathtt{P}_{\rm c}$. This is critical, given the fact that batteryless RFID tags typically incorporate no energy storage element, e.g., (super)capacitor, due to size and cost limitations.
%\subsection{Tag Powering}
%\label{subsec:tag_powering}

\subsection{Backscatter Communication}
\label{subsec:backscatter_comm}
As stated earlier, the tag alters the load terminating its antenna using a switch. Load $Z_0$ is, by construction,
designed to match antennas' impedance. Thus, when antenna is terminated at $Z_0$, the load absorbs (ideally, 
if perfectly matched) all the power offered by the impinged signal. When the antenna is terminated at $Z_1$, a 
fraction $\rho_{\rm u} \leq  1- \tau_{\rm d}$ of the impinged power is used for uplink scatter radio operation. 
Parameter $\rho_{\rm u}$ depends on the tag scattering efficiency (which also incorporates non-idealities  from the above model).
Modified reflection coefficient\cite{BlDiSa:10} $\Gamma_{i}$, when the antenna is terminated at $Z_i, ~i  \in \{0,1\}$, 
is given by ${\Gamma}_{i} = \frac{ {Z}_i - \mathtt{Z}_{\rm a}^{*}}{ {Z}_i + \mathtt{Z}_{\rm a}}$, where $\mathtt{Z}_{\rm a}$ 
antenna's impedance. The baseband equivalent of the tag-backscattered signal can be expressed as \cite{BlDiSa:10} 
$\mathtt{A}_s - \Gamma_{i}$, which in turn depends on the (load-independent) tag antenna structural mode $\mathtt{A}_s$ 
and the transmitted bit $i$; the backscattered baseband signal, for a duration of $N$ tag bits, is given by \cite{KiBlSi:13_2}:
%\vspace{-15pt}
\begin{align}
\hspace{-5pt}
 \mathsf{b}(t) = &\sqrt{\mathsf{L}  \rho_{\rm u}  \mathtt{P}_{\rm R}} h    \Big(\mathtt{A}_s - \Gamma_0 + \Delta\Gamma \sum_{n=1}^{N}  \mathsf{s}_{b_n}\!(t-(n-1)T)\Big),
\end{align}
where, $\Delta\Gamma \triangleq (\Gamma_0 - \Gamma_1)$, $b_n \in \{0,1\}$ is the $n$-th reflected bit,  while  function $\mathsf{s}_{b_n}(\cdot)$ is the  backscattered signal
basis function, of duration $T$, when bit $b_n$ is transmitted.

In order to a) balance the time for which the tag is absorbing energy, 
independently of the tag's data bits, and b) avoid \emph{ghost} tag reception, 
i.e., reader misinterpreting thermal noise as tag information, a \emph{line} 
code is used in commercial GEN2 RFIDs \cite{rfid_epc:08}, selecting between 
FM0 and Miller. Under FM0 coding, observing $2T$ signal duration for each bit (of duration $T$) suffices for BER-optimal, coherent (differential) detection and
$\mathsf{s}_{b_n}\!(\cdot)$ is a $T/2$-shifted waveform given by \cite{KaMaBl:15}:
\begin{align}
\mathsf{s}_{0}(t) \triangleq  
 \begin{cases}
 1 , &   0 \le t < \frac{T}{2}, \\
 0 ,  &    \text{otherwise},
 \end{cases}  ~
 \mathsf{s}_{1}(t) \triangleq  
 \begin{cases}
 1 , &  \frac{T}{2}\le t < T, \\
 0 ,  &    \text{otherwise}.
 \end{cases}
\end{align}
%{\color{red} Incorporating fading}, the received (at the reader) backscattered signal becomes:
%\begin{equation}
% \widetilde{\mathsf{y}}(t) = \sqrt{\mathsf{L}} \, h \, \mathsf{b}(t) + \mathsf{n}(t),
%\end{equation}
%where  $\mathsf{n}(t)$ is a  circularly-symmetric  complex   Gaussian noise  process. Some of the waveforms $\{\mathsf{s}_{b_n}(\cdot)\}_{n=1}^N$ are known to the reader for channel estimation and synchronization purposes.

%Using \cite[Theorem 1]{FaAlBl:15}, the baseband signal at the output of correlators for a duration of $N$ symbols becomes:

Assuming perfect synchronization, the optimal demodulator projects the received signal onto the basis functions subspace 
using two correlators. The discrete baseband signal, at the output of the correlators, follows \cite[Theorem 1]{FaAlBl:15}:
\begin{align}
 \mathbf{y}_n  = g \,\mathbf{s}_n + \mathbf{w}_n,  ~n=1,2,\ldots, N,
 \label{eq:baseband_signal_model}
\end{align}
where 
$ g \triangleq  \mathsf{L} \sqrt{\rho_{\rm u}  \, \mathtt{P}_{\rm R}} \, h^2   
 \, (\Gamma_0 - \Gamma_1),
$
and $\mathbf{s}_n$ is the vector representation for the $n$-th
transmitted signal. For RFID systems, which employ $T/2$-shifted FM$0$ line-coding, $\mathbf{s}_n \in \left\{ [1~~ 0]^{\top}, [0~~ 1]^{\top}\right\}$
and $\mathbf{w}_n \sim \mathcal{CN}(\mathbf{0}_2, \sigma^2 \, \mathbf{I}_2)$  \cite{FaAlBl:15, KaMaBl:15}, with $\sigma^2$   denoting the variance of each noise component.
 
%while vector $ \mathbf{w}_n$ is circularly-symmetric complex Gaussian vector with diagonal covariance matrix. 

%; in other words, tag information is modulated on the (complex) reflection coefficient of the system tag antenna-tag connected input load. 
%tag scatters back the incident signal. 
%Describe backscatter operation. Compressed backscatter operation text.

\section{Reader}
\label{sec:reader}
%Reader related stuff. Outage scenarios. Successful reception probability.
%For the baseband signal in Eq.~\eqref{eq:baseband_signal_model}, given known channel $g$, and coherent ML differential detection, the conditional bit error probability (BER) for RFID systems employing FM0 line coding, follows from \cite{KaMaBl:15, SimDiv:06}:

%\begin{equation}
% \mathsf{Q}(x) \approx \frac{1}{2}\, \mathsf{e}^{-0.374x^2 - 0.777x}, \, x\geq 0,
%\end{equation}

\subsection{Bit Error Rate (BER)}
\label{subsec:rfid_ber}
Assuming coherent ML differential detection (with signal of $2T$ duration, given known channel $g$), the conditional bit error probability for the baseband 
signal in Eq.~\eqref{eq:baseband_signal_model} follows from \cite{KaMaBl:15, SimDiv:06}:
\begin{equation}
 \mathbb{P}({\rm error }|  g) =  
  2 \,\mathsf{Q}\!\left(\frac{|g|}{\sigma}\right) \left(  1 - \mathsf{Q}\!\left(\frac{|g|}{\sigma}\right) \right),
  \label{eq:BER_ML_conditional_1}
\end{equation}
where $\mathsf{Q}(x) =\frac{1}{\sqrt{2 \pi}} \int_x^{\infty} \mathsf{e}^{-\frac{t^2}{2}} \mathsf{d}t $ is the  Q-function.
Interestingly, a similar expression applies to Miller line coding, when the receiver performs coherent (ML) bit-by-bit 
detection.

\subsection{Outage Scenarios}
\label{subsec:outage_scenarios}
%Describe it outage scenario: Havesting Sensitivity Outage, Power Consumption Outage and Information Outage (ber under fixed/defined-threshold)

The reader receives successfully the RFID tag's information when: a)  the input RF power at the tag antenna is above RF harvesting sensitivity, and b) the harvested power is above tag's power consumption, given that the RFID tag does not include energy storage elements, and c) BER at
the reader is below a threshold $\beta$. Probability of these events is analyzed below.

\subsubsection{Outage due to limited harvesters' sensitivity}
Considering the definition of input power in~Eq.~\eqref{eq:input_power}, 
tag's harvesting sensitivity outage metric is defined as follows:
\begin{equation}
\mathbb{P}(  \mathscr{A}) \triangleq \mathbb{P}( {P}_{\rm in} \leq \mathtt{P}_{\rm sen}) =  \mathsf{F}_{P_{\rm in}}(\mathtt{P}_{\rm sen}), 
 \label{eq:sens_outage_event}
\end{equation}
where $\mathsf{F}_{P_{\rm in}}(\cdot)$ is the cumulative distribution function (CDF) of $P_{\rm in}$. Eq.~\eqref{eq:sens_outage_event} mathematically describes the probability that the input power ${P}_{\rm in}$ at the RFID 
tag antenna (which depends on the wireless channel/fading), is below tag RF harvester's sensitivity $\mathtt{P}_{\rm sen}$. Such outage event represents the fraction of time the tag's rectenna cannot harvest RF energy due to inadequate input RF power. Under Nakagami fading such outage is given by:
\begin{equation}
\mathsf{F}_{P_{\rm in}}(\mathtt{P}_{\rm sen})  = 1 - \int_{\mathtt{P}_{\rm sen}}^{\infty} 
\mathsf{f}_{P_{\rm in}} (y)  \mathsf{d} y = 1- 
\frac{\mathsf{\Gamma}\!  \left(  \mathtt{M},  \frac{ \mathtt{M}}{ \mathsf{L} \,\mathtt{P}_{\rm R}} \mathtt{P}_{\rm sen} \right)}{\mathsf{\Gamma}(\mathtt{M})} ,
\end{equation}
where   $\mathsf{\Gamma}\!  \left(  \alpha ,  z \right) =  \int_{z}^\infty  t^{\alpha-1} \mathsf{e}^{-t} \mathsf{d}t$. At this point, it must be emphasized that RF receiver sensitivity for communication purposes can obtain values from $-80$ dBm or less, while state-of-the-art rectennas offer harvesting sensitivity in the order of around $-40$ to $-35$ dBm \cite{AsDaBle:16}.
Clearly, \emph{signals useful for communication may not be useful for power transfer}.

\begin{figure}[!t]
\centering
        \includegraphics[width=0.76\columnwidth]{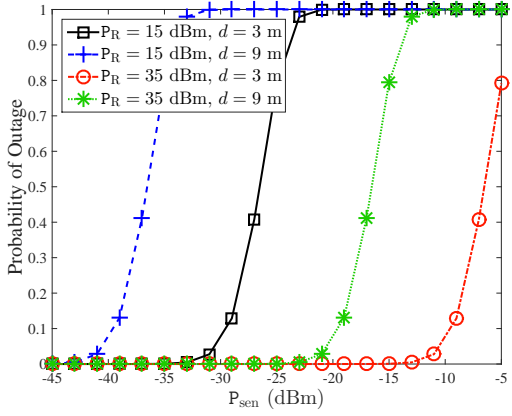}
\caption{Probability of sensitivity outage event as a function of tag's harvesting sensitivity. The path-loss model of Eq.~\eqref{eq:path_loss} is employed with  $\nu = 2.1$, $\lambda =0.3456$ and $\mathtt{M}=5$.}
\label{fig:impact_sensitivity}
\end{figure}
Fig.~\ref{fig:impact_sensitivity} examines Eq.~\eqref{eq:sens_outage_event} as a function of  tag RF harvester's sensitivity. It can be clearly 
seen that less-sensitive RF harvesters, suffer from higher outage probabilities. RF harvesting sensitivity is commonly neglected in SWIPT
research, even though it tremendously impacts the \emph{power transfer} part and thus, overall performance \cite{AlTouBle:18}.
%
%Two harvesters from prior art are further discussed. The first RF harvester is the commercial PowerCast module  \cite{PowerCast}, with offered sensitivity of $\mathtt{P}_{\rm sen} = -12$ dBm; the probability of sensitivity outage is almost $1$ for transmission power $P_{ \rm R} = 20$ dBm and distance $d$ more than $4$ meters. The second RF harvester is the sensitive rectenna in \cite{AsDaBle:16} with $\mathtt{P}_{\rm sen} = -42.5$ dBm, thus, the outage event becomes almost $0$ for all studied parameters   $P_{ \rm R}$  and $d$. RF harvesting sensitivity is commonly neglected in SWIPT research, even though it tremendously impacts the \emph{power transfer} part and thus, overall performance.

\subsubsection{Outage due to limited power consumption}
When input power is above tag's harvesting sensitivity, the next type of outage is when the harvested power, $\mathsf{p}( \zeta_{\rm har}  \,  P_{\rm in} )$ is \emph{not enough}, i.e., below tag's power consumption $\mathtt{P}_{\rm c}$:
\begin{equation}
\mathbb{P}( \mathsf{p}( \zeta_{\rm har}  \,  P_{\rm in} ) \leq \mathtt{P}_{\rm c}), 
\label{eq:lim_harv_power_outage_event}
\end{equation}
which depends on (a) fading and input power at the tag, (b) the type of the RF harvester, and (c) tag's power consumption $\mathtt{P}_{\rm c}$; such probability describes the fraction of time the harvested power is not adequate for tag powering and is critical for 
devices that cannot store harvested energy. If $\mathsf{p}(\cdot)$ is strictly increasing and continuous around $\mathtt{P}_{\rm c}$  \cite{Fol:99}, the event  in Eq.~\eqref{eq:lim_harv_power_outage_event} can be simplified as follows:
\begin{equation}
\mathbb{P}( \mathscr{B}) \triangleq   \mathbb{P}\!\left(    P_{\rm in}   \leq \frac{\mathsf{p}^{-1}(\mathtt{P}_{\rm c})}{\zeta_{\rm har}} \right)  =
\mathsf{F}_{P_{\rm in}}\!\!\left( \frac{\mathsf{p}^{-1}(\mathtt{P}_{\rm c})}{\zeta_{\rm har}} \right), 
\label{eq:lim_harv_power_outage_event_v2}
\end{equation}
where  $\mathsf{p}^{-1}(\mathtt{P}_{\rm c})$ is the inverse function of $\mathsf{p}(\cdot)$ at point $\mathtt{P}_{\rm c}$.

\subsubsection{Information Outage}
RFID tag information outage at the reader is defined when BER in Eq.~\eqref{eq:BER_ML_conditional_1} 
 is below a predefined precision $\beta$. Setting $\mathsf{R}(x) \triangleq    2 \,\mathsf{Q}( x ) \, (1 -  \mathsf{Q}(x))  , ~x \in  (0,\infty) $, this event can be mathematically expressed as \cite{AlBl:18}:
 \begin{align}
 \mathbb{P}( \mathscr{C}) \triangleq 
   \mathbb{P}\!\left( P_{\rm in} \leq \frac{  \sqrt{\mathtt{P}_{\rm R}} \,\sigma  \, \mathsf{R}^{-1}(\beta)} {|\Gamma_0 - \Gamma_1| \sqrt{\rho_{\rm u}}}\right)
   = \mathsf{F}_{P_{\rm in}}\!\!\left(\frac{  \sqrt{\mathtt{P}_{\rm R}} \,\sigma  \, \mathsf{R}^{-1}(\beta)} {|\Gamma_0 - \Gamma_1| \sqrt{\rho_{\rm u}}} \right),  
 \label{eq:info_outage}
\end{align}
where $\mathsf{R}^{-1}(x) =  \mathsf{Q}^{-1}\!\left( \frac{1 - \sqrt{1 - 2\,x}}{2} \right) $, defined for $x \in (0,0.5)$
and $ \mathsf{Q}^{-1}(\cdot)$ is the inverse of Q-function.

\subsection{Probability Of Successful Reception}
\label{subsec:prob_success}
Tag information is unsuccessfully received when \emph{either} of previously discussed events $\mathscr{A}$, $\mathscr{B}$, $\mathscr{C}$ occurs. 
Assuming that function $\mathsf{p}(\cdot)$ is strictly increasing and continuous around $\mathtt{P}_{\rm c}$ and denoting for an event 
$\mathscr{D}$ its complement as $\mathscr{D}^{\mathtt{C}}$,
the probability of  unsuccessful SWIPT reception, denoted as event $\mathscr{F}$,  can be expressed as:
\begin{align}
 \mathbb{P}(\mathscr{F}) & =1-\mathbb{P}(\mathscr{F}^{\mathtt{C}}) =1 -\mathbb{P}( \mathscr{A}^{\mathtt{C}} \cap \mathscr{B}^{\mathtt{C}} \cap \mathscr{C}^{\mathtt{C}}) \nonumber \\  
  &=  1 - \mathbb{P}\!\left(   P_{\rm in}  >  \theta_{\mathscr{F}} \right) = \mathsf{F}_{ P_{\rm in}}\!\!\left( \theta_{\mathscr{F}}\right),   
  \label{eq:prob_succ_reception_reader_abstract}
 \end{align}
where $\theta_{\mathscr{F}} \triangleq \max\left\{\mathtt{P}_{\rm sen},\frac{\mathsf{p}^{-1}(\mathtt{P}_{\rm c})}{\zeta_{\rm har}}, \frac{ \sqrt{\mathtt{P}_{\rm R}} \,\sigma \, \mathsf{R}^{-1}(\beta)} {|\Gamma_0 - \Gamma_1|\, \sqrt{\rho_{\rm u}}} \right\}$. Consequently,  successful SWIPT reception at the reader,  under Nakagami fading, is given in closed form as follows:
\vspace{-6pt}
\begin{align}
\label{eq:succ_reception}
\mathbb{P}({\rm SWIPT ~success}) \equiv \mathbb{P}(\mathscr{F}^{\mathtt{C}}) =
\frac{\mathsf{\Gamma}\!  \left(  \mathtt{M}, \frac{ \mathtt{M}}{ \mathsf{L} \,\mathtt{P}_{\rm R}} \theta_{\mathscr{F}}  \right)}{\mathsf{\Gamma}(\mathtt{M})}.
\end{align}

\section{Numerical Results}
\label{subsec:num_results}

%$J+1 = 118$ data points (input power-harvested power) were used.
%To study the probability of successful SWIPT reception at the reader of RF-powered tag, path-loss
For the simulation results the path-loss model of Eq.~\eqref{eq:path_loss} is considered with $\nu = 2.3$ and $\lambda =0.3456$ (UHF carrier frequency), and 
tag antenna reflection coefficients $\Gamma_0$ and $\Gamma_1$ satisfying $|\Gamma_0 - \Gamma_1| = 1$.  
The ultra-sensitive harvester in \cite{AsDaBle:16} is tested using parameters   $\tau_{\rm d} = 0.5$, 
$\chi = 0.5$, $\rho_{\rm u}= 0.01$  for RF harvesting and backscattering at the tag, while BER threshold is set
$\beta = 10^{-5}$; variance of noise at the reader was set to $10^{-11}$.

%%%BER PLOT & TEXT
%%%  \begin{figure}[!t]
%%%  \centering
%%%          \includegraphics[width=0.82\columnwidth]{Figures/BER_vs_Ptx_v1.eps}   
%%%   \caption{Probability of bit error as a function of reader's transmit power.}  
%%%  \label{fig:BER_vs_Ptx_dist}
%%%  \end{figure}
%%%   
%%%   {
%%%  \color{red}
%%%   First, Fig.~\ref{fig:BER_vs_Ptx_dist} studies the impact of reader's transmit power on BER 
%%% assuming Nakagami fading with $\mathtt{M}=6$ and two different distance setups $d=5$ m and $d = 8$ m.
%%% It can be seen that the proposed BER approximation in~\eqref{eq:final_BER_approx} is in full accordance with
%%%     Monte Carlo  simulation. It can be observed that for ${\rm BER} = 0.01\%$, $23$ and $32$ dBm transmit 
%%% power is required for tag-to-reader distance $d = 5$ and $d = 8$, respectively.
%%% }

\begin{figure}[!t]
\centering
        \includegraphics[width=0.75\columnwidth]{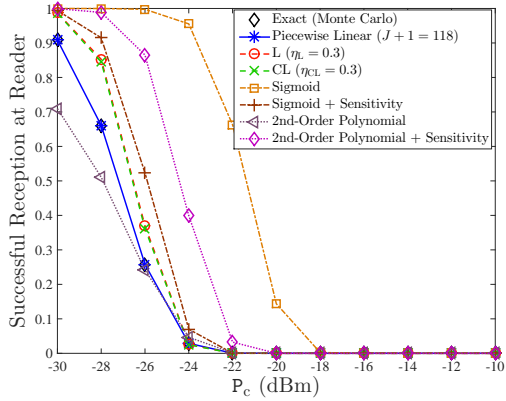}
\caption{Probability of successful SWIPT reception at reader, as a function of tag power consumption-Strong LoS.}
\label{fig:probSucc_strongLoS}
\end{figure}

\begin{figure}[!t]
\centering
        \includegraphics[width=0.75\columnwidth]{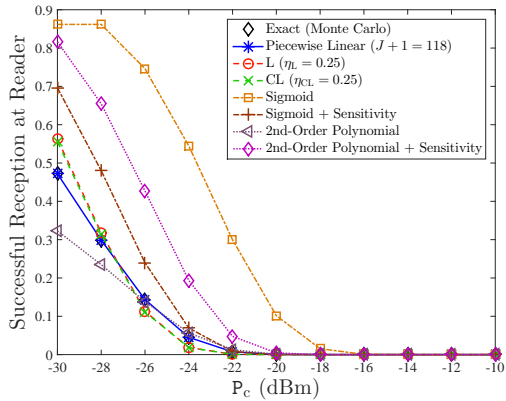}
\caption{Probability of successful SWIPT reception at reader, as a function of tag power consumption-non LoS.}
\label{fig:probSucc_nonLoS}
\end{figure}

Fig.~\ref{fig:probSucc_strongLoS} depicts probability of successful SWIPT reception at the reader, as a function of tag's power consumption, 
in a strong LoS scenario (Nakagami parameter $\mathtt{M}=10$), $d = 4$ m, and $P_{\rm R} = 1$ Watt. Fig.~\ref{fig:probSucc_nonLoS} examines
the same relationship in a non-LoS scenario ($\mathtt{M}=2$), $d = 7$ m, and $P_{\rm R} = 2.5$ Watt.

Both figures clearly show that the performance of the piecewise linear model $\mathsf{p}_7(\cdot)$ coincides with the exact  (ground-truth, 
${\mathsf{p}}(  \cdot )$), data-driven model. The performance of $\mathsf{p}_1(\cdot)$ (L), as well as $\mathsf{p}_2(\cdot)$ 
(CL) model deviate from reality, even though the best values for the efficiency parameters  were utilized (i.e., values 
that offered performance as close as possible to the ground-truth model). 
Both nonlinear sigmoid models tend to overestimate the event while the one incorporating sensitivity, offers closer-to-reality results in the
LoS scenario and deviates further in the non-LoS scenario. Finally, the second-order polynomial $\mathsf{p}_5(\cdot)$ underestimates performance, with performance gap that depends on the scenario and
tag's power consumption, whereas  energy harvesting model  $\mathsf{p}_6(\cdot)$ overestimates the harvested power. 
In short, SWIPT research requires \emph{accurate} energy harvesting models, otherwise misleading conclusions are unavoidable.

%.
\vspace{-5pt}
\section{Conclusion}
\label{sec:conclude}
SWIPT research should always take into account all the non-ideal characteristics of the RF energy harvesting system; otherwise, oversimplification due to overlooking fundamentals from electronics and microwave engineering may lead to impractical results. This work studied the sensitivity and the nonlinearity of the harvester. Impact of other modules, present in the RF harvesting chain (e.g., boost converter/maximum power point tracking-MPPT), should be also examined.
\vspace{-5pt}
\section*{Acknowledgment}
This research is implemented through the Operational Program ``Human Resources Development, Education and Lifelong Learning'' and is co-financed by the European Union (European Social Fund) and Greek national funds.

%It is evident that SWIPT research is interdisciplinary and should be always carefully 
%coupled with fundamentals in the electronics and microwave community; otherwise, relevant
%protocol and algorithmic design may be impractical.
%We touched upon the sensitivity and nonlinearity 
%of the harvester. Other important building blocks, such as the boost converter/maximum power point tracking 
%(MPPT) module's impact, important in any energy harvesting method, should be also examined.

\bibliographystyle{IEEEtran}
\bibliography{IEEEabrv,BLETSAS_GROUP_bib_v13}

\begin{thebibliography}{10}
\providecommand{\url}[1]{#1}
\csname url@samestyle\endcsname
\providecommand{\newblock}{\relax}
\providecommand{\bibinfo}[2]{#2}
\providecommand{\BIBentrySTDinterwordspacing}{\spaceskip=0pt\relax}
\providecommand{\BIBentryALTinterwordstretchfactor}{4}
\providecommand{\BIBentryALTinterwordspacing}{\spaceskip=\fontdimen2\font plus
\BIBentryALTinterwordstretchfactor\fontdimen3\font minus
  \fontdimen4\font\relax}
\providecommand{\BIBforeignlanguage}[2]{{%
\expandafter\ifx\csname l@#1\endcsname\relax
\typeout{** WARNING: IEEEtran.bst: No hyphenation pattern has been}%
\typeout{** loaded for the language `#1'. Using the pattern for}%
\typeout{** the default language instead.}%
\else
\language=\csname l@#1\endcsname
\fi
#2}}
\providecommand{\BIBdecl}{\relax}
\BIBdecl

\bibitem{Du:16}
G.~D. Durgin, ``{RF} thermoelectric generation for passive {RFID},'' in
  \emph{Proc. IEEE RFID}, Orlando, FL, May 2016, pp. 1--8.

\bibitem{Ale_PhD:17}
P.~Alevizos, ``Intelligent scatter radio, {RF} harvesting analysis, and
  resource allocation for ultra-low-power internet-of-things,'' Ph.D.
  dissertation, School of ECE, Technical University of Crete, Chania, Greece,
  2017, advisor: A. Bletsas.

\bibitem{AlBl:18}
\BIBentryALTinterwordspacing
P.~N. Alevizos and A.~Bletsas, ``Sensitive and nonlinear far field {RF} energy
  harvesting in wireless communications,'' \emph{{IEEE} Trans. Wireless
  Commun.}, 2018, accepted, to appear. [Online]. Available:
  \url{https://arxiv.org/pdf/1707.07041.pdf}
\BIBentrySTDinterwordspacing

\bibitem{CleBaz:16}
B.~Clerckx and E.~Bayguzina, ``Waveform design for wireless power transfer,''
  \emph{{IEEE} Trans. Signal Process.}, vol.~64, no.~23, pp. 6313--6328, Dec.
  2016.

\bibitem{Cler:18}
B.~Clerckx, ``Wireless information and power transfer: Nonlinearity, waveform
  design and rate-energy tradeoff,'' \emph{{IEEE} Trans. Signal Process.},
  vol.~66, no.~4, pp. 847--862, Feb. 2018.

\bibitem{AsDaBle:16}
S.~D. {Assimonis}, S.-N. {Daskalakis}, and A.~{Bletsas}, ``Sensitive and
  efficient {RF} harvesting supply for batteryless backscatter sensor
  networks,'' \emph{{IEEE} Trans. Microw. Theory Techn.}, vol.~64, no.~4, pp.
  1327--1338, Apr. 2016.

\bibitem{Goldsmith:05}
A.~Goldsmith, \emph{Wireless Communications}.\hskip 1em plus 0.5em minus
  0.4em\relax New York, NY, USA: Cambridge University Press, 2005.

\bibitem{BoNgZlSc:15}
E.~Boshkovska, D.~W.~K. Ng, N.~Zlatanov, and R.~Schober, ``Practical non-linear
  energy harvesting model and resource allocation for {SWIPT} systems,''
  \emph{{IEEE} Commun. Lett.}, vol.~19, no.~12, pp. 2082--2085, Dec. 2015.

\bibitem{WaXiHuWu:17}
S.~Wang, M.~Xia, K.~Huang, and Y.~C. Wu, ``Wirelessly powered two-way
  communication with nonlinear energy harvesting model: Rate regions under
  fixed and mobile relay,'' \emph{{IEEE} Trans. Wireless Commun.}, vol.~16,
  no.~12, pp. 8190--8204, Dec. 2017.

\bibitem{XuOzKeVi:17}
X.~Xu, A.~\"Ozcelikkale, T.~McKelvey, and M.~Viberg, ``Simultaneous information
  and power transfer under a non-linear {RF} energy harvesting model,'' in
  \emph{Proc. IEEE Int. Conf. Communications}, Paris, France, May 2017, pp.
  179--184.

\bibitem{GuKleCheYaAMi:09}
N.~J. Guilar, T.~J. Kleeburg, A.~Chen, D.~R. Yankelevich, and R.~Amirtharajah,
  ``Integrated solar energy harvesting and storage,'' \emph{{IEEE} Trans.
  {VLSI} Syst.}, vol.~17, no.~5, pp. 627--637, May 2009.

\bibitem{FoGuKlePhaYaAmi:13}
E.~G. Fong, N.~J. Guilar, T.~J. Kleeburg, H.~Pham, D.~R. Yankelevich, and
  R.~Amirtharajah, ``Integrated energy-harvesting photodiodes with diffractive
  storage capacitance,'' \emph{{IEEE} Trans. {VLSI} Syst.}, vol.~21, no.~3, pp.
  486--497, Mar. 2013.

\bibitem{BlDiSa:10}
A.~Bletsas, A.~G. Dimitriou, and J.~Sahalos, ``Improving backscatter radio tag
  efficiency,'' \emph{{IEEE} Trans. Microw. Theory Techn.}, vol.~58, no.~6, pp.
  1502 -- 1509, Jun. 2010.

\bibitem{KiBlSi:13_2}
J.~Kimionis, A.~Bletsas, and J.~N. Sahalos, ``Bistatic backscatter radio for
  power-limited sensor networks,'' in \emph{Proc. IEEE Global Commun. Conf.
  ({G}lobecom)}, Atlanta, GA, Dec. 2013, pp. 353--358.

\bibitem{rfid_epc:08}
\emph{{EPC} Radio-Frequency Identity Protocols, Class-1 Generation-2 {UHF}
  {RFID} Protocol for Communications at 860 {MHZ}-960 {MHZ}}.\hskip 1em plus
  0.5em minus 0.4em\relax {EPC} Global, 2008, version 1.2.0.

\bibitem{KaMaBl:15}
N.~Kargas, F.~Mavromatis, and A.~Bletsas, ``Fully-coherent reader with
  commodity {SDR} for {G}en2 {FM0} and computational {RFID},'' \emph{{IEEE}
  Wireless Commun. Lett.}, vol.~4, no.~6, pp. 617--620, Dec. 2015.

\bibitem{FaAlBl:15}
N.~Fasarakis-Hilliard, P.~N. Alevizos, and A.~Bletsas, ``Coherent detection and
  channel coding for bistatic scatter radio sensor networking,'' \emph{{IEEE}
  Trans. Commun.}, vol.~63, pp. 1798--1810, May 2015.

\bibitem{SimDiv:06}
M.~Simon and D.~Divsalar, ``Some interesting observations for certain line
  codes with application to {RFID},'' \emph{{IEEE} Trans. Commun.}, vol.~54,
  no.~4, pp. 583--586, Apr. 2006.

\bibitem{AlTouBle:18}
\BIBentryALTinterwordspacing
P.~N. Alevizos, K.~Tountas, and A.~Bletsas, ``Multistatic scatter radio sensor
  networks for extended coverage.'' [Online]. Available:
  \url{arxiv.org/pdf/1706.03091.pdf}
\BIBentrySTDinterwordspacing

\bibitem{Fol:99}
G.~B. Folland, \emph{Real analysis: Modern techniques and their applications},
  2nd~ed.\hskip 1em plus 0.5em minus 0.4em\relax John Wiley \& Sons, Inc., New
  York, 1999.

\end{thebibliography}

%figures
\end{document}